\documentclass[12pt]{spie}  
\usepackage{amsmath,amsfonts,amssymb}
\usepackage{graphicx}
\usepackage{setspace}
\usepackage{tocloft}
\usepackage{lineno}
\usepackage{authblk} 
\usepackage{url}

\newcommand\arcsec{\mbox{$^{\prime\prime}$}}%
\newcommand\farcs{\mbox{$.\!\!^{\prime\prime}$}}%

\title{Technology and Science Advancing Observations with Roman Coronagraph Informed by Ground-Based High-Contrast Imaging}

\author[1]{Mona El Morsy}
\author[1,2]{Thayne Currie}
\author[3]{Brianna Lacy}
\author[1]{Danielle Bovie}
\author[1]{Erica Dykes}
\author[1]{Jie Li}
\author[2,4,5,6]{Olivier Guyon}
\author[2,4]{Julien Lozi}
\author[2,4]{Garima Singh}
\author[2,7]{Kyoohun Ahn}
\author[2,8]{Vincent Deo}
\author[2,9,10]{Sebastien Vievard}
\author[2]{Yoshito Ono}
\affil[1]{Department of Physics and Astronomy, University of Texas at San Antonio, San Antonio, TX 78006, USA}
\affil[2]{National Astronomical Observatory of Japan, Subaru Telescope, 650 North Aohoku Place, Hilo, HI 96720, USA}
\affil[3]{NASA Ames Research Center, Moffett Field, CA, USA}
\affil[4]{Astrobiology Center, 2-21-1, Osawa, Mitaka, Tokyo 181-8588, Japan}
\affil[5]{Steward Observatory, University of Arizona, Tucson, AZ 85721, USA}
\affil[6]{College of Optical Sciences, University of Arizona, Tucson, AZ 85721, USA}
\affil[7]{Korea Astronomy and Space Science Institute (KASI), Daejeon 34055, Republic of Korea}
\affil[8]{Optical Sharpeners, Manosque, France}
\affil[9]{Space Science and Engineering Initiative, College of Engineering, University of Hawai‘i, Hilo, HI 96720, USA}
\affil[10]{Institute for Astronomy, University of Hawaii, Hilo, HI 96720, USA}

\cftpagenumbersoff{figure}
\cftpagenumbersoff{table} 
\begin{document} 
\maketitle

\begin{abstract}
The Roman Coronagraph technology demonstration focuses on achieving $<$ 10$^{-7}$ contrasts within the instrument's dark hole and our ability to detect and characterize properties of faint companions around bright stars. Here, we describe results from a study of potential Roman Coronagraph technology demonstration phase observations focused on these goals, informed by the ongoing OASIS survey at the Subaru Telescope and precursor survey work. OASIS provides at least three compelling targets for the technology demonstration phase with imaged companions – the HIP 71618 B brown dwarf and superjovian planets HIP 54515 b and HIP 99770 b.   HIP 71618 is well suited for demonstrating the Coronagraph’s core performance requirement while all three targets are well suited for spectroscopic mode observations.  Each target can be paired with a PSF reference star vetted for companions.   While HIP 71618 and HIP 54515 are already planned for Technology Demonstration phase observations, we describe the programmatic and scientific value of adding spectroscopic mode observations of HIP 99770 as well.

\end{abstract}



{\noindent \footnotesize\textbf{Send correspondence to M. El Morsy},  \url{mona.elmorsy@utsa.edu} }


\section{Introduction}
Direct imaging is a means that will someday characterize the atmospheres and orbits of terrestrial zone rocky planets around nearby stars \cite{Currie2023b}.  Over the past decade, \textit{extreme} adaptive optics (AO) facilities such as SPHERE, GPI, and SCExAO have matured this technique from the ground \cite{Chauvin2017,Macintosh2015,Keppler2018,Currie2022,Currie2023a}, but they largely focus on detecting young, self-luminous superjovian planets in the IR, where thermal emission and favorable planet-to-star contrasts (e.g. 10$^{-4}$--10$^{-6}$) render companions detectable.  Dedicated extreme AO systems on the upcoming \textit{extremely large telescopes} should yield contrasts capable of imaging rocky, habitable zone planets around M stars in reflected light ($\sim$10$^{-8}$; see \cite{Kasper2021}).  But imaging 10$^{-9}$--10$^{-10}$ contrast Jupiter to Earth twins around Sun-like stars is best achieved in a highly stable space-borne environment.

The Coronagraph Instrument (CGI) aboard the Nancy Grace Roman Space Telescope will be the first space-based instrument capable of reflected-light planet detections.  CGI employs deformable mirrors for high-order wavefront sensing and control and advanced starlight suppression to carve out a high-contrast dark hole around target stars \cite{Kasdin2020}. As a technology demonstration, CGI's core performance requirement -- Threshold Technical Requirement 5 (TTR5) -- is to reach a contrast of 10$^{-7}$ at $\sim$ 6--9 $\lambda$/D from a V $\sim$ 5 host at $\lambda$ $<$ 600 nm \cite{Bailey2023}.
Beyond TTR5, CGI is tasked with fulfilling five separate ``Objectives" (2.2.1--2.2.5), many of which are related to the ability to do exoplanet science at deep optical contrasts from space.   E.g. Objective 2.2.5 requires characterizing ``\textit{photometry, spectroscopy, and astrometry"} of at least one companion.  

Mature planets in reflected light have contrasts too steep for detectability at the TTR5-level contrast of 10$^{-7}$.  However, warm (T $\sim$ 1000-3000 K) self-luminous planets and brown dwarfs detected by ground-based extreme AO systems can have optical contrasts that require CGI's instrumentation but are easier to detect than mature planets \cite{LacyBurrows2020,Bovie2025, ElMorsy2024b,ElMorsy2025,Currie2026a}.  The $>$42 night OASIS survey (PI. T. Currie, Co-PI M. Kuzuhara) was funded by the NASA Science Mission Directorate specifically to find such targets \cite{ElMorsy2024a}. 

In this work, we evaluate OASIS-studied systems as candidate targets for Roman Coronagraph technology-demonstration-phase observations aimed at achieving TTR5 and fulfilling CGI's science Objectives, building upon analysis briefly presented in several Roman Coronagraph White Papers \cite{Currie2026b,Currie2025a,Currie2026d}. We focus on three targets with imaged companions — the brown dwarf HIP 71618 B and the superjovian planets HIP 54515 b and HIP 99770 b — and assess their suitability for demonstrating the core contrast requirement and Objective 2.2.5 requiring spectroscopic characterization. For each science target, we further vet selected candidate PSF reference stars. Section \ref{sec:sample} describes the targets and summarizes the current literature consensus on their classification.  Section \ref{sec:analysis} predicts the companions' locations during the Roman Coronagraph technology demonstration phase, estimates their contrasts in the Roman Coronagraph passbands, presents preliminary binary companion screening results for PSF reference stars, and assesses scheduling feasibility.  

For all analysis, we assume a launch date of August 30, 2026 followed by a commissioning phase that lasts until November 30, 2026.

\section{Sample of Roman Coronagraph Targets}\label{sec:sample}

\begin{figure}
\begin{center}
\includegraphics[width=1\textwidth,clip]{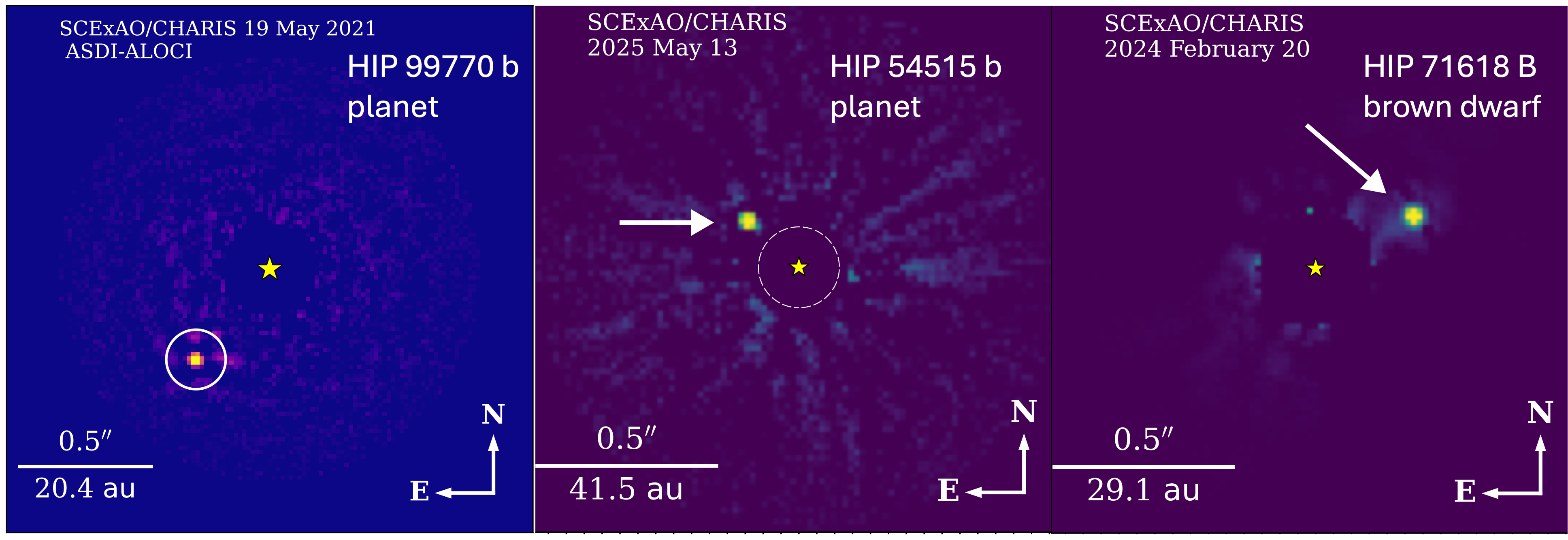}
\end{center}
\vspace{-0.25in}
\caption 
{ \label{fig:sample}
Discovery images of targets in our sample: (left) the HIP 99770 b planet \cite{Currie2023a}, (middle) the HIP 54515 b planet \cite{Currie2026a}, and the HIP 71618 B companion (likely a brown dwarf; \cite{ElMorsy2025}). } 
\end{figure} 

Figure \ref{fig:sample} displays annotated discovery images of our sample companions.  We detail their properties below:

\begin{itemize}
    \item \textbf{HIP 99770 b} is a recently discovered but now well-studied directly-imaged superjovian planet\cite{Currie2023a} orbiting the nearby ($\sim$40 pc) bright (V = 4.9) A5-A6 star HIP 99770 A (29 Cyg, HD 192640).  While the star's kinematics are consistent with membership in the 40 Myr-old Argus association, the system likely has an age comparable to the Pleiades or AB Doradus (115-200 Myr; see Supplementary Material in \cite{Currie2023a}).   The current consensus is that it is the first joint direct imaging and astrometric discovery of an extrasolar planet.   
    
    Dynamical modeling favors a companion mass of $\sim$ 13--15 $M_{\rm Jup}$, a semimajor axis of $\sim$ 17 au, and an eccentricity of $e$ $\sim$ 0.3--0.4 \cite{Currie2023b,Bovie2025,Winterhalder2025,Balmer2026}.  Atmospheric modeling suggests that the planet is a late L dwarf with a cloudiness and surface gravity intermediate between that of the HR 8799 planets and field brown dwarfs \cite{Bovie2025}.  The planet's atmosphere is enhanced in metals and its orbit is aligned with the spin axis of the star at the 2-$\sigma$ level.  Combined together, these features are strongly suggestive of a bottom-up, core accretion-like formation in a protoplanetary disk and confirm the planet interpretation of this companion \cite{Balmer2026}.   HIP 99770 b joins HR 8799 bcde as one of the few planets whose formation is well probed by its atmospheric composition \cite{Xuan2026}.

    \item \textbf{HIP 54515 b} is a superjovian planet discovered from the OASIS survey \cite{Currie2026a} orbiting a more distant (83 pc) A2V star.  The system's age is less well constrained, with likely values between 20 Myr to 200 Myr.  Dynamical modeling favors a companion mass of $\sim$ 17--18 $M_{\rm Jup}$, a semimajor axis of $\sim$25 au, and an eccentricity of 0.4. Its CHARIS spectrum is best matched by an object at the M/L transition.


    \item \textbf{HIP 71618 B} is a low-mass companion to the young A2V star HIP 71618 also discovered from the OASIS survey \cite{ElMorsy2025}.  The system is also young although its age is not well constrained (20--200 Myr).   Its mass posterior drawn from dynamical modeling covers 60$^{+27}_{-21}$ $M_{\rm Jup}$ or 65$^{+54}_{-29}$ $M_{\rm Jup}$, depending on the assumed prior.  
    The companion is likely on a high eccentricity orbit with a semimajor axis of $\approx$ 11 au.  We interpret it as a \textit{likely} brown dwarf (see below).

\end{itemize}

   Both HIP 99770 b and HIP 54515 b likely have masses at or slightly exceeding the classical deuterium-burning limit (DBL), estimated to be $\approx$ 13 $M_{\rm Jup}$ for solar helium abundances, where a substellar companion will consume 50\% of its deuterium over Gyr timescales\cite{Spiegel2011}.  While the DBL was once considered to be a plausible dividing line between planets and brown dwarfs \cite{Lecavelier2022}, this criterion was simply declared a working definition by fiat and has been decisively refuted in the peer-reviewed literature\footnote{As noted in the HIP 99770 b discovery paper and later work \cite{Currie2023b,Currie2026a}, the DBL itself was also very early on criticized as being internally incoherent as a criterion -- e.g. see Luhman \cite{Luhman2008} -- while theory papers explicitly cast doubt on whether it is meaningful either \cite{Spiegel2011}.  None of these criticisms have been confronted in the literature, let alone undermined.}.  A crossover between planets and brown dwarfs at $\approx$25 $M_{\rm Jup}$ and/or a mass ratio of $q$ $\sim$ 0.025 and separation of $a$ $\lesssim$ 100-300 au is best justified \cite{Currie2023a,Currie2026a,Giacalone2026}, though the mass ratio cutoff may be more fundamental than mass \cite{Zhang2025}.  Recent results from JWST/NIRCam indicate that companions this massive can show evidence for a core accretion-like formation in a protoplanetary disk \cite{Balmer2026}.  Thus, both HIP 99770 b and HIP 54515 b are properly considered planets based on current evidence.

    For HIP 71618 B, most of the mass posterior distribution lies below the standard threshold -- $\approx$75--80 $M_{\rm Jup}$ -- dividing objects that have a hydrogen-burning main sequence from those that do not\footnote{Unlike deuterium burning, hydrogen burning \textit{does} have a substantial impact on an object's long-term evolution and thus does constitute a meaningful discriminator between different types of astrophysical objects.}.   The exact division between stellar and substellar, though, may be dependent on metallicity and clouds, ranging between $\approx$ 66 $M_{\rm Jup}$ and 82 $M_{\rm Jup}$ \cite{Morley2024}.  Thus, we consider it to be a \textit{likely} brown dwarf.

\section{Analysis}
\label{sec:analysis}

\subsection{Angular Separation During the Roman Technology Demonstration}
\begin{figure}
\begin{center}
\includegraphics[width=1\textwidth,clip]{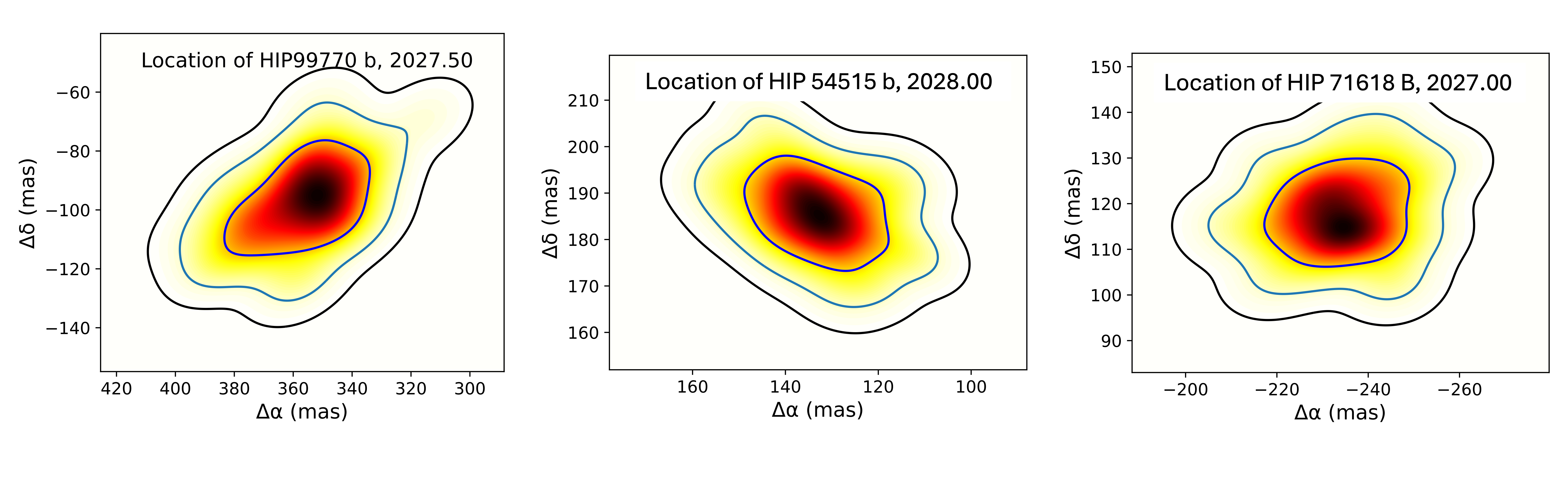}
\end{center}
\vspace{-0.25in}
\caption 
{ \label{fig:location}
Predicted locations of HIP 99770 b, HIP 54515 b, and HIP 71618 B during the early stages of the Roman Coronagraph technology demonstration phase drawn from \texttt{orvara} dynamical modeling results presented in \cite{Currie2023a,Bovie2025,Currie2026a,ElMorsy2025}.}
\end{figure}
To predict each companion's location during the Roman Coronagraph technology demonstration phase, we jointly model their relative astrometry (from imaging) with absolute astrometry of the star from Hipparcos and Gaia eDR3 using \texttt{orvara} \cite{Brandt2021}.  The \texttt{orvara} code samples the full orbital posterior while simultaneously constraining the companion's dynamical mass. Propagating these posteriors forward in time yields the companions' predicted on-sky position during the demonstration phase window. 

Figure~\ref{fig:location} shows the resulting relative-position ($\Delta\alpha~,
\Delta\delta$
) probability distributions for HIP 99770 b at epoch 2027.5, HIP 54515 b at 2028.0, and HIP 71618 B at 2027.0, with the nested contours enclosing the 68\%, 95\%, and 99.7\% credible regions. The sampled epochs effectively assume the following: that HIP 71618 B will be the first post-commissioning target attempted (to fulfill TTR5) while HIP 54515 b and HIP 99770 b will be imaged later, with scheduling interleaved with other Coronagraph and non-Coronagraph observations (e.g. reflected light imaging of $\epsilon$ Eridani; the Roman High-Latitude Time-Domain Survey).

In each case, the companion's predicted location is constrained to within a few tens of mas.  During the first year of the Technology Demonstration Phase, the companions will be located at projected separations of $\approx$0\farcs25--0\farcs4, placing them within CGI's dark hole at 575 nm and 730 nm. 
At slightly earlier and later epochs, HIP 99770 b and HIP 54515 b retain roughly the same angular separation but only change in position angle.  HIP 71618 B lies on an edge-on orbit.  For the earliest feasible post-commissioning epoch ($\sim$ Dec 1 2026), we predict an angular separation of 0\farcs{}26.   HIP 71618 B will stay within the Coronagraph's dark hole region through at least the first part of 2028.   

\subsection{Predicted Contrast in the Roman Coronagraph Passbands
}
\begin{figure}
\begin{center}
\includegraphics[width=1\textwidth,clip]{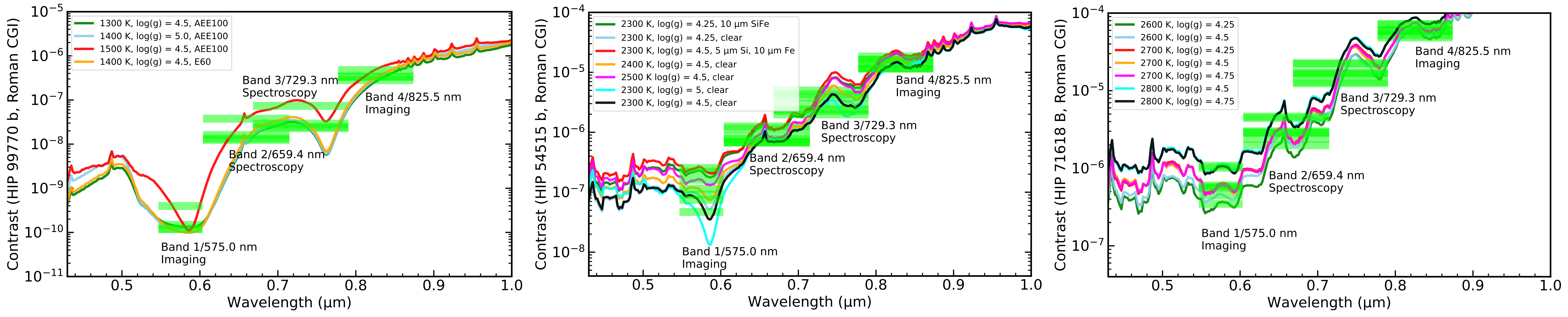}
\end{center}
\vspace{-0.25in}
\caption 
{ \label{fig:contrast}
Predicted contrast of HIP 99770 b, HIP 54515 b, and HIP 71618 B in the Roman Coronagraph major passbands for photometry and spectroscopy, drawing from the atmospheric modeling presented in \cite{Bovie2025}, during the early stages of the Roman Coronagraph technology demonstration phase drawn from \texttt{orvara} dynamical modeling results presented in \cite{Currie2023a,Bovie2025,Currie2026a,ElMorsy2025}.}
\end{figure}
Next, we predict each companion's brightness in the Roman Coronagraph passbands.  We adopt predictions from the \cite{LacyBurrows2020} substellar atmosphere models, given best-fitting atmospheric parameters for HIP 99770 b from Bovie et al \cite{Bovie2025}, for HIP 54515 from Currie \& Li et al \cite{Currie2026a}, and HIP 71618 B from El Morsy et al \cite{ElMorsy2025}.  We compute the companions' optical flux densities in the Coronagraph passbands, assuming a flat transmission profile over the published bandpass limits. 

Figure \ref{fig:contrast} shows our predictions.  Because these are warm, self-luminous objects, the companions' predicted contrasts are substantially larger than those predicted based on the reflected light component alone ($\sim$10$^{-10}$--10$^{-12}$). In the 575 nm bandpass, HIP 71618 B has a predicted contrast of $\sim$3--10$\times$10$^{-7}$, a result corroborated by predictions based on empirical optical spectra of mid M dwarfs (see \cite{ElMorsy2025}).   Thus, a high signal-to-noise ratio detection of HIP 71618 B (e.g. SNR $\sim$ 50) will achieve TTR5.  HIP 54515 b is slightly cooler and thus its 575 nm contrast is slightly steeper: 7$\times$10$^{-8}$ to 2.5$\times$10$^{-7}$.   While a SNR $\gtrsim$ 12.5 detection of HIP 54515 b would also demonstrate TTR5, the primary is significantly fainter than the preferred TTR5 target brightness (V $\lesssim$ 5).  HIP 99770 b's predicted 575 nm contrast is significantly steeper ($\lesssim$ 10$^{-9}$): the planet is likely undetectable with Roman at 575 nm.

In the 730 nm passband for Coronagraph spectroscopy, all three companions are detectable.  At a contrast of $\sim$10$^{-5}$, HIP 71618 B should provide a clear demonstration of extracting atmospheric properties in the high SNR limit for bright stars.   Provided that the Roman Coronagraph still achieves deep contrast ($\lesssim$ 10$^{-7}$) on V = 6.8 stars, HIP 54515 b's detection at $\sim$10$^{-6}$ contrast should likewise be decisive and further demonstrate spectral extraction at small angular separations.   Finally, HIP 99770 b's predicted contrast of 10$^{-7}$--10$^{-8}$ lies slightly below the TTR5 threshold but within the predicted performance of the instrument.  

\subsection{PSF Reference Star Vetting}
The Roman Coronagraph's observing strategy requires that each science target is paired with a bright PSF reference star for dark hole digging and reference PSF subtraction.  Suitable PSF reference stars must be brighter than V $\sim$ 3, have an angular size $\lesssim$2 mas, must lack substantial circumstellar dust, and must lack companions capable of compromising CGI's DH digging at TTR5-relevant levels ($\gtrsim$ 10$^{-7}$--10$^{-8}$ contrast) \cite{Hom2026}.   The Roman Community Participation Program (CPP) defined a list of candidate reference stars for Band 1 (575 nm) observations with the Hybrid Lyot coronagraph and other modes\footnote{\url{https://docs.google.com/spreadsheets/d/1p5r0VmjBCjXU25daJl5oJOPoPh1V79nuESbnwmca0s0/edit?gid=1677191164\#gid=1677191164}}, ranked as ``A", ``B", or ``C" based on these criteria listed.  

Using a combination of optical interferometry, speckle interferometry, and shallow AO imaging, the Roman CPP team provided an initial vetting of reference stars.   While they ruled out stellar companions bright in the near-IR, the corresponding contrast limits in $V$ band are typically $\gtrsim$ 10$^{-4}$.  These contrasts are a factor of 1000 brighter than the limit needed to vet the references for TTR5 suitability.    

Therefore, we obtained SCExAO/CHARIS high-contrast imaging of candidate PSF reference stars in April and May 2026.  We focus on northern reference stars located in or near the Roman Continuous Viewing Zone (CVZ): regions more than 54$^{o}$ from the ecliptic plane that are the focus of the Roman High-Latitude Time-Domain Survey \cite{Rose2023}.   
PSF reference stars and tech demo targets within or near the CVZ maximize the schedulability of Coronagraph observations, as their observations can be more flexibly slotted in during gaps in the Survey cadences.  
Of the four candidate reference stars observed, three are drawn from the CPP reference star list: $\alpha$ Cep, $\epsilon$ UMa, and $\eta$ UMa. A fourth star, $\gamma$ Boo, was identified in \cite{ElMorsy2025} but is not included in the current CPP reference star list, likely because its apparent V-band magnitude is $\sim$ 3.03.
We reduced the data with the CHARIS Data Processing Pipeline \cite{Currie2020b}, performing basic reduction steps (sky subtraction, image registration, spectrophotometric calibration, spatial filtering, etc.) and PSF subtraction using A-LOCI \cite{Currie2012,Currie2015}.   

\begin{figure}
\begin{center}
\includegraphics[width=1\textwidth,clip]{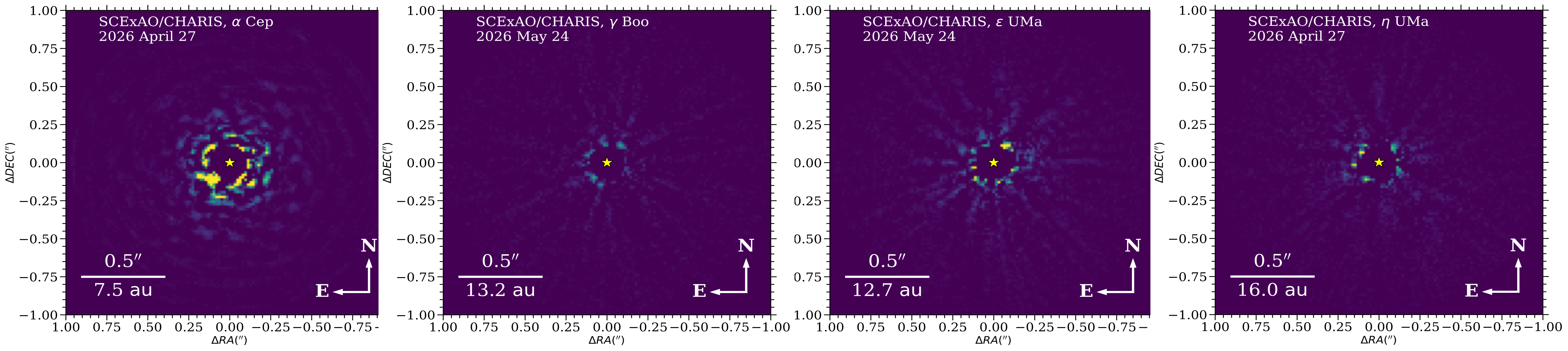}
\end{center}
\vspace{-0.25in}
\caption 
{ \label{fig:psfref}
Sequence-combined SCExAO/CHARIS images of candidate Roman Coronagraph PSF reference stars obtained with SCExAO/CHARIS.  In each case, we performed PSF subtraction with ADI only \cite{Marois2006}.  Incorporating SDI \cite{SparksFord2002} yields a contrast improvement between a factor of $\sim$ 1.25 to 2.5 between 0\farcs{}2 and 1\arcsec{}: results from such reductions will be presented in a later paper.  }
\end{figure}

Figure \ref{fig:psfref} shows reductions of these data sets with PSF subtraction combined with \textit{angular differential imaging}\cite{Marois2006} only.  We fail to detect any residuals consistent with a companion at the 5-$\sigma$ significance level.  For most sources, point source contrast limits with ADI are $\sim$10$^{-5}$ at 0\farcs{}25--0\farcs{}3 and 3$\times$10$^{-6}$ at 0\farcs{}5, although limits for $\alpha$ Cep are worse due to low-wind effect \cite{Milli2018}.  Limits using both ADI and \textit{spectral differential imaging} (SDI)\cite{SparksFord2002} and a more thorough analysis will be presented in an upcoming paper.  

\subsection{Roman Coronagraph Scheduling}

\begin{figure}
\begin{center}
\includegraphics[width=1\textwidth,trim=0mm 0mm 0mm 0mm,clip]{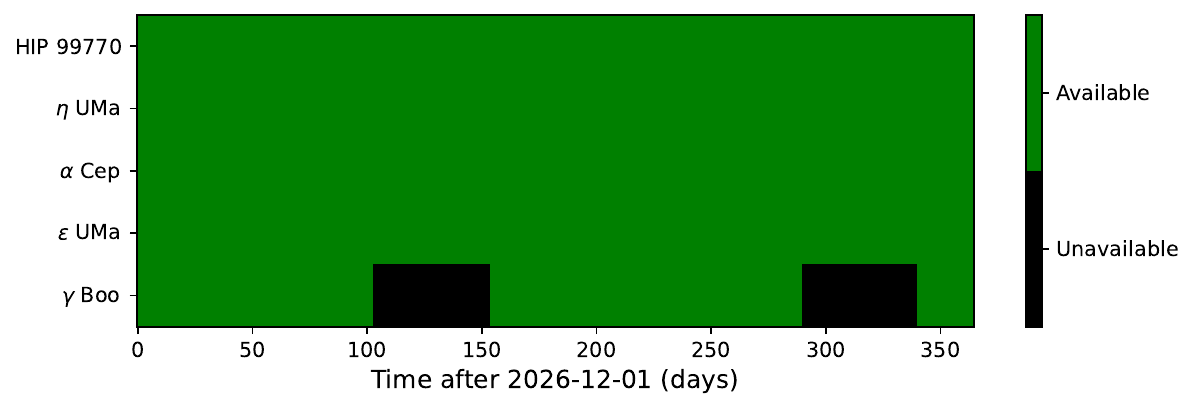}\\
\includegraphics[width=0.4\textwidth,trim=0mm 0mm 0mm 0mm,clip]{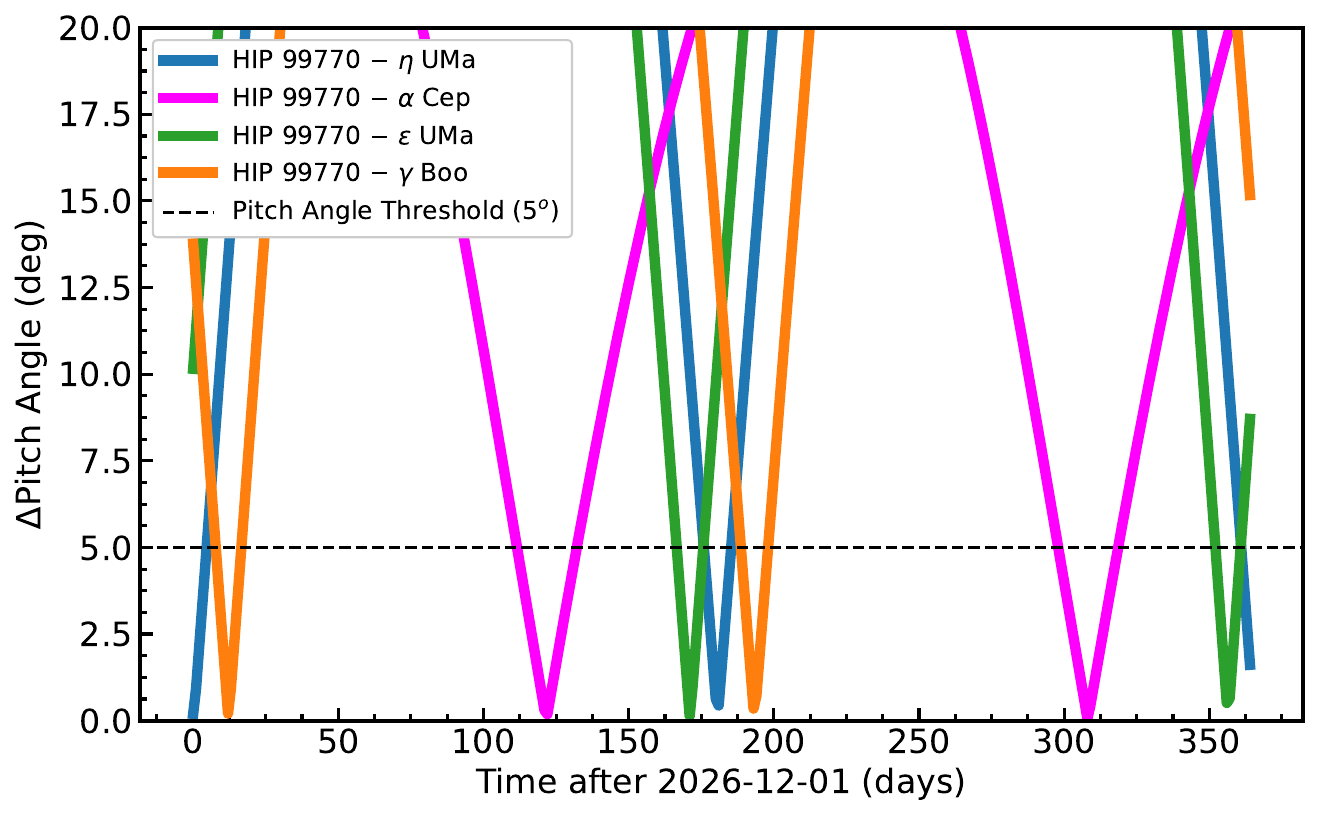}
\includegraphics[width=0.425\textwidth,trim=300mm 0mm 0mm 0mm,clip]{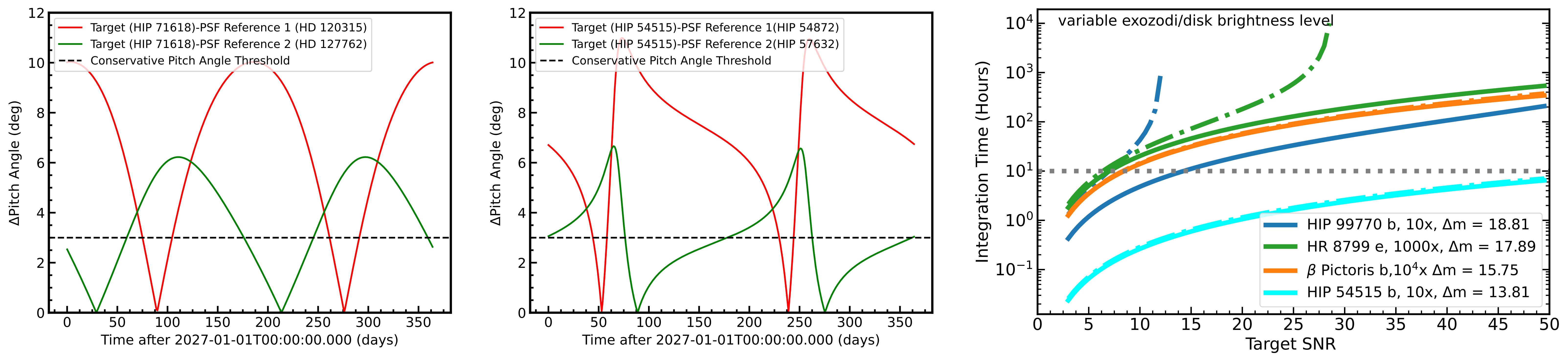}
\end{center}
\vspace{-0.25in}
\caption 
{ \label{fig:schedule}
(top) Keep-out map and (bottom-left) delta-pitch angle plot for HIP 99770 observations paired with our PSF reference stars.  (bottom-right) 
Exposure time estimate to achieve a SNR = 10 detection at 730nm for HIP 99770 b and HIP 54515 b planet compared to other planets.  The horizontal dotted line denotes 10 hours, a nominal observing time limit for technology demonstration targets.}
\end{figure}

Technology Demonstration Phase observations must also satisfy Roman's pointing and thermal constraints, which limit the range of solar pitch angles accessible at any given epoch.  Reference stars impose an additional, tighter constraint.  The PSF reference and science target must have a nearly identical thermal environment, to within $\sim$ 5$^{o}$, otherwise the dark hole generated with the reference will degrade.  Thus, we compute the pitch-angle difference between each target and its vetted reference as a function of observing epoch.

Figure \ref{fig:schedule} shows example scheduling analysis for HIP 99770 paired with PSF reference star observations. Based purely on its solar angle, HIP 99770 is well observable throughout the year, while the PSF reference stars are either observable for 75\% of the year ($\gamma$ Boo) or also over the entire year ($\alpha$ Cep, $\epsilon$ UMa, $\eta$ UMa).   Pitch angle requirements make schedulable time more restrictive, although HIP 99770 observations are possible with these PSF reference stars over $\sim$ 1/3 of each calendar year: two 60-day blocks extending from $\sim$ November to December and May to June.  The Roman Coronagraph exposure time calculator suggests that spectroscopic mode observations at 730nm should yield a characteristic SNR of $\sim$10 within 10 hours of on-source time.   Similar analysis for HIP 54515 and HIP 71618 demonstrates that these targets are also easily schedulable (see \cite{ElMorsy2025,Currie2026a,Currie2026b,Currie2026d}).

\section{Discussion}
We have analyzed three OASIS-discovered systems with imaged companions - the likely brown dwarf HIP 71618 B and the superjovian planets HIP 54515 b and HIP 99770 b - for suitability as targets for Roman Coronagraph technology-demonstration-phase observations. All three targets' companions will lie within the deep-contrast dark hole region for the Roman Coronagraph during the technology demonstration phase.  At 575 nm, HIP 71618 B lies at contrast just higher than the TTR5 threshold of 10$^{-7}$, while HIP 54515 is at or slightly below the TTR5 threshold. Both companions have very modest (10$^{-5}$--10$^{-6}$) contrasts on the 730 nm spectroscopic band; HIP 99770 b is also likely detectable with 730 nm spectroscopy (10$^{-8}$--10$^{-7}$).  All three targets can be paired with vetted PSF reference stars over substantial windows within the first year of operations.

Since we began this analysis, the CPP has assembled a preliminary technology-demonstration target list\footnote{\url{http://conference.ipac.caltech.edu/SpiritofLyot6/slides/Wednesday\_Morning\_Wolff\%20-\%20Schuyler\%20Wolff.pdf}}.  HIP 71618 is among the first post-commissioning targets scheduled and will be used to demonstrate TTR5 in Band 1.  It and HIP 54515 are likewise planned for spectroscopic mode observations.

While not currently on the provisional schedule, HIP 99770 warrants substantial consideration.   It is predicted to be about a factor of 2 fainter at 730 nm than HR 8799 e, the canonical warm-planet spectroscopy target.  Because of its brightness ($V$ $\sim$4.9 vs. $\sim$5.96 for HR 8799) and excellent year-round visibility, HIP 99770 may be an easier target to achieve the Coronagraph's best contrast and schedule efficiently.  Both HIP 99770  and HR 8799 have outer cold, Kuiper belt-like disk components.  However, HIP 99770 has no known evidence for a warm dust component that could confuse the planetary companion's signal\cite{Currie2023a}, while HR 8799 includes a luminous warm dust belt that could be a potential contaminant \cite{Boccaletti2024}.  Consequently, depending on the system's dust properties, HR 8799 e may be more challenging to detect (see Figure \ref{fig:schedule}, lower-right panel).  

HIP 99770 b spectra coupled with those of HR 8799 e are also scientifically valuable.  While the two planets have comparable spectral types and temperatures (late L; $\sim$1100-1200 K vs. $\sim$1300 K), HR 8799 e is very cloudy, while HIP 99770 b has a cloudiness intermediate between those of the youngest late L dwarf planets like HR 8799 e and field brown dwarfs with similar temperatures \cite{Currie2011,Bovie2025, Balmer2026}, making it arguably a more tractable object to interpret.  The 730nm bandpass probes the potassium doublet feature (0.77 $\mu m$), which is dependent on both metallicity and clouds.  Scheduling both planets for 730nm observations therefore enables robust comparative exoplanet spectroscopy probing key atmospheric features.

\acknowledgements
Based on data collected at Subaru Telescope, which is operated by the National Astronomical Observatory of Japan. 

The development of SCExAO and AO3k is supported by the Japan Society for the Promotion of Science (Grant-in-Aid for Research \#23340051, \#26220704, \#23103002, \#19H00703, \#19H00695 and \#21H04998), the Subaru Telescope, the National Astronomical Observatory of Japan, the Astrobiology Center of the National Institutes of Natural Sciences, Japan, the Mt Cuba Foundation and the Heising-Simons Foundation. The development of the CACAO software is supported by the National Science Foundation under award 2410616. CHARIS was built at Princeton University under a Grant-in-Aid for Scientific Research on Innovative Areas from MEXT of the Japanese government (\#23103002). The authors wish to recognize and acknowledge the very significant cultural role and reverence that the summit of Maunakea has always had within the Hawaiian community, and are most fortunate to have the opportunity to conduct observations from this mountain. 

The authors declare that there are no financial interests, commercial affiliations, or other potential conflicts of interest that could have influenced the objectivity of this research or the writing of this paper.


\bibliography{bibliography}   
\bibliographystyle{spiebib}   





\end{document}